\documentclass[12pt]{article}
\usepackage{amsmath}
\usepackage{color}
\usepackage{epsfig}
\usepackage{graphicx}
\usepackage{scrtime}
\usepackage[multiple]{footmisc}

\usepackage{amsfonts, graphics}

\begin{document}

\newcommand{\IC}{\mathbb C}
\newcommand{\IR}{\mathbb R}
\newcommand{\IQ}{\mathbb Q}
\newcommand{\IZ}{\mathbb Z}
\newcommand{\IN}{\mathbb N}
\newcommand{\mk}{\marginpar}
\newcommand{\Rn}{\IR^n}
\newcommand{\be}{\begin{enumerate}}
\newcommand{\ee}{\end{enumerate}}
\newcommand{\dl}{\displaystyle}

\title{Cosmic evolution with a general Gaussian type scale factor}
\author{Subenoy Chakraborty   \footnote{schakraborty@math.jdvu.ac.in}  \\ Department of Mathematics, Jadavpur University, Kolkata 700 032, India.         \and
Subhra Bhattacharya \footnote{subhra.maths@presiuniv.ac.in} \\
Department of Mathematics, Presidency University, Kolkata 700 073, India.        
}
\vspace{1em}

\maketitle
\hrulefill

\vspace{2em}
{\large\bf Abstract:} Here we present a simple model of cosmic evolution in Einstein gravity, with the cosmic substratum being composed of an inhomogeneous and anisotropic fluid. The scale factor is supposed to be of Gaussian type. In this framework we show the existence of a continuously evolving eternal universe with no singularity, beginning or end.

\hrulefill

\vspace{1em}

In this essay we will discuss about a model of cosmic evolution based on the generalization of the emergent universe scenario. It is known that following Einstein’s discovery of general theory of relativity (GTR) since 1917 there has been a humongous amount of developments in terms of cosmology in general that has led to the evolution of some fascinating and testable theories of the universe. Modern day scientists predict an expanding universe which has been now established by using several observational evidences emerging from galaxy red-shift surveys and cosmic microwave background (CMB) \cite{teg}, \cite{kow}, \cite{kom}, \cite{plank1}, \cite{plank2}. The most compelling models based on GTR and cosmological principles of homogeneity and isotropy of such an expanding universe is the Big Bang model that predicts that the universe evolves from a time like singularity. But initial time like singularity essentially means a point in time where physical laws breakdown. Alternative cosmological models providing singularity free cosmic evolution are of bouncing type cosmologies \cite{bounce} and the emergent universe models \cite{harri}. Emergent universe models provide examples for past eternal Einstein static inflationary universes. If developed carefully emergent universes can solve the conceptual problem associated with initial time like singularity in contemporary models. However most of these models are incomplete in the sense that they provide explanations to some specific observable or physics predicted in the course of cosmic evolution. The bouncing universe models on the other hand are cyclical or oscillatory models of the universe where the first cosmological event is interpreted as the result of a collapse of a previous universe. However bouncing theories failed due to their incompatibility with inflationary universes.

We, therefore try to contemplate a scenario where space and time existed for ever and evolved continuously. Our motivation is to suggest an alternative evolutionary theory of the universe that is continuous and ever existing. With this aim we start our description of a spherically symmetric model of the universe having matter component that is anisotropic and inhomogeneous. The underlying theory being guided by Einstein’s GTR without any time like strong singularity.
The line element for such an universe is assumed to be 
\begin{equation}
ds^{2}=-dt^{2}+a(t)^{2}\left[\frac{dr^{2}}{1-b(r)}+r^{2}(d\theta^{2}+\sin^{2}\theta d\phi^{2})\right].\label{metric}
\end{equation}
Here $a(t)$ is the scale factor and $b(r)$ is some function of the radial coordinate $r.$ Discussion on the nature of $b(r)$ is kept for later. We perceived that the properties of the scale factor can be of some importance in actuating the notion of a non-singular universe. As such scale factor is one dynamical variable that is present in every aspect of universe's evolution and can reflect upon the main events in the history and future of the cosmic evolution. Further we wanted a scale factor that can mix the qualities of both emergent and bouncing cosmologies. Heuristically we therefore decide upon the following generalized Gaussian distribution as our scale factor:
\begin{equation}
a=a_{0}+a_{1}e^{-\mu(t-t_{0})^{2}}\label{a}
\end{equation} with $a_{0},~a_{1},\mbox{and}~\mu$ all arbitrary positive constants. We observe that the above choice of the scale factor does not support the existence of any strong singularity of type 1, 2 or 3 \cite{odi}. In fact existence of any of the above singularities can lead to a physically unrealistic situation which is classically considered to be a flaw in a theory. 

The matter making up our universe is assumed to be inhomogeneous and anisotropic having $(\rho, p_{r},p_{t})$ as the corresponding energy density and anisotropic pressures. Considering the Einstein's field equations for the above metric, we find that fluid satisfies the usual Friedmann's equations given by,

\begin{equation}
\left.\begin{aligned}
\kappa\rho(t,r)=3H^{2}+\frac{b(r)+rb'(r)}{a^{2}r^{2}}\\
\kappa p_{r}=-2\dot{H}-3H^{2}-\frac{b(r)}{a^{2}r^{2}}\\ 
\kappa p_{t}=-2\dot{H}-3H^{2}-\frac{b'(r)}{2a^{2}r}
\end{aligned}
\right\}
\label{f2}
\end{equation} with $\kappa=8\pi G,$ the gravitational coupling parameter. They satisfy the corresponding energy conservation equations:
\begin{align}
\frac{\partial\rho}{\partial t}+H(3\rho+p_{r}+2p_{t})=0  \\
\frac{\partial p_{r}}{\partial r}=-\frac{2}{r}(p_{r}-p_{t}).\label{ec1}
\end{align}

It is evident that using the above $a(t)$ the energy density and thermodynamic pressure for the fluid can be exactly computed using the given expressions. Further an analytical evaluation of the mathematical expression for the Hubble parameter and acceleration parameter help us conclude the following: 

That the universe expands during $(-\infty,t_{0})$ and contracts in $(t_{0},\infty).$ During its expanding era, it evolves initially in an accelerating manner in the interval $(-\infty,t_{1})$ while expansion is decelerating in the interval $(t_{1},t_{0}),$ with $t_{1}=t_{0}-\frac{1}{\sqrt{2\mu}}.$ At $t=t_{0}$ expanding universe undergoes phase shift and begins contracting. Initially contraction occurs in a decelerating manner till $t=t_{2}=t_{0}+\frac{1}{\sqrt{2\mu}}.$ From $(t_{2},\infty)$ contraction is accelerated. Asymptotically as $t\rightarrow\pm\infty$ we find that the Hubble parameter, scale factor and corresponding matter densities are given by:
\begin{align*}
\text{i)}~ &H\rightarrow 0,~a\rightarrow a_{0}\\
\text{ii)}~ &\kappa\rho\rightarrow\frac{b(r)+rb'(r)}{a_{0}^{2}r^{2}},~\kappa p_{r}\rightarrow-\frac{b(r)}{a_{0}^{2}r^{2}},~\kappa p_{t}\rightarrow-\frac{b'(r)}{2a_{0}^{2}r}.
\end{align*}
Accordingly we can conclude that the above modelled universe was in Einstein static phase in infinite past and then after cosmic evolution it will again come back to its emergent state in future infinity. 

To give our model a firmer basis we search for inflationary regimes in our model. Classically the rate of inflation roll is defined using two parameters $\epsilon_{H}$ and $\eta_{H}.$ Slowly varying Hubble parameter or accelerated expansion corresponds to $0<\epsilon_{H}<1$ where $\epsilon_{H}$ the first slow roll parameter is defined as $-\dot{H}/H^{2}.$ Considering inflation of the slow roll type, required the second slow roll parameter $\eta_{H}=\frac{\dot{\epsilon_{H}}}{H\epsilon_{H}}<<1.$ Recent studies have revealed that similar results can be obtained by considering a constant rate of inflation roll where the second slow roll parameter is such that $\eta_{H}$ is constant or finite \cite{cr}. Our model will successfully compare to constant roll inflationary regimes near the two asymptotic ends $t=\pm\infty.$ We have
\begin{equation}
\epsilon_{H}=\frac{a_{0}e^{\mu(t-t_{0})^{2}}[1-2\mu(t-t_{0})^{2}]-a_{1}}{2\mu a_{1}(t-t_{0})^{2}}
\end{equation} 
The requirement $0<\epsilon_{H}<1$ gives 
\begin{equation}
-1<-2\sqrt{e}W(x)\left(\frac{a_{0}}{a_{1}}\right)<x
\end{equation}
where $x=2\mu(t-t_{0})^{2}-1$ and $W(x)=\frac{x}{2}e^{\frac{x}{2}}$ is the Lambert's $W$ function. For real $x,~W(x)$ is real for the following two cases:
\begin{itemize}
\item[] $-\frac{1}{e}\leq W(x)<0\Rightarrow -1\leq \frac{x}{2}<0\Rightarrow t_{0}-\frac{1}{\sqrt{2\mu}}<t<t_{0}+\frac{1}{\sqrt{2\mu}}.$ This gives $-\frac{1}{2\sqrt{e}p}<\left(\frac{a_{0}}{a_{1}}\right)<-\frac{2\delta}{2\sqrt{e}p},$ where $\frac{x}{2}=-\delta$ and $W(x)=-p$, both $p$ and $\delta$ being positive. Since by choice both $a_{0}$ and $a_{1}$ are constrained to be positive, this relation can never be satisfied and hence $\epsilon_{H}<1$ shall not be achieved in this case.

\item[]$W(x)>0\Rightarrow \frac{x}{2}>0\Rightarrow t>t_{0}+\frac{1}{\sqrt{2\mu}}$ and $t<t_{0}-\frac{1}{\sqrt{2\mu}}$ which gives $0<\left(\frac{a_{0}}{a_{1}}\right)<\frac{1}{2\sqrt{e}p}.$ Putting the stated restriction on the parameters $a_{0}$ and $a_{1}$ we can make $\epsilon_{H} <1$ in the two regions around $t=(-\infty,t_{1})$ and $t=(t_{2},\infty).$ 
\end{itemize}
Correspondingly the parameter $\eta$ is given by:
$\eta_{H}=\left(\frac{a_{0}}{a_{1}}\right)\frac{2e}{1-2p\sqrt{e}(a_{0}/a_{1})}\left[\frac{\delta}{p\sqrt{e}}\left(1+\frac{\delta}{2}\right)+\frac{a_{0}}{a_{1}}\left(1+2\delta-\sqrt{\frac{1+2\delta}{2\mu}}\right)\right]$ is a finite quantity. Thus our model gives two inflationary regimes of constant roll type that will exist at both the asymptotic ends. The first inflationary regime: Inflation $I$ between  $-\infty<t<t_{1}$ will signify accelerated expansion while the second inflationary regime: Inflation $II$  at $t_{2}<t<\infty$  will signify accelerated contraction. It is interesting to note that using the Gaussian scale factor it is possible to exhibit two sides of inflation, namely both expansion and contraction. Further inflationary expansion channels an expanding universe while contracting inflation channels the same universe into shrinking universe keeping the cycle of alternate expansion and contraction in motion. 

Next we answer the question how the contracting past universe is channelled into an expanding present universe. At $t=\pm\infty$ with the choice of $b(r)=b_{0}r^{2}+\frac{b_{1}}{r},~b_{0}$ and $b_{1}$ being arbitrary constants, the metric is one that describes a traversable wormhole \cite{m+t}, \cite{visserb} given by
\begin{equation}
ds^{2}=-dt^{2}+\frac{d\zeta^{2}}{1-\frac{B(\zeta)}{\zeta}}+\zeta^{2}(d\theta^{2}+\sin^{2}\theta d\phi^{2}).\label{whmetric}
\end{equation} 
Here $\zeta=a_{0}r$ and $\frac{B(\zeta)}{\zeta}=B_{0}\zeta^{2}+\frac{B_{1}}{\zeta},~B_{0}=\frac{b_{0}}{a_{0}^{2}}$ and $B_{1}=b_{1}a_{0}.$ The above wormhole is characterised by throat at $\zeta_{0}=B_{1}$ and $B_{0}$ corresponding to the curvature constant $K$ and can take values $\pm 1$ or 0. With wormhole configurations we know that energy conditions and their violations are important. In this model the null energy condition (NEC) is satisfied for $\left(\frac{B(\zeta)}{\zeta}\right)'>0$ while NEC is violated for $\left(\frac{B(\zeta)}{\zeta}\right)'<-2B(\zeta)/\zeta^{2}.$ Hence we hypothesise this static wormhole as the bridge (or tunnel) connecting the two Einstein static scenarios at the two asymptotic regions ({\it i.e} past and future infinity).

Analysing the stability of such a static wormhole configuration using the cut-paste technique of the usual Israel-Darmois formalism we arrive at the following relations (by matching the interior static wormhole geometry with a similar static exterior geometry):
\begin{equation}
\sigma'\eta|_{\alpha_{0}}\geq F\left(\frac{B(\alpha_{0})}{\alpha_{0}}\right).
\end{equation}
Where $\eta=\frac{P'}{\sigma'},$ with $\sigma$ and $P$ the surface stress energy and pressure at the junction interface and $'$ indicates differentiation w.r.t. $\alpha$. Here $\alpha(\tau)>\zeta_{0}$ being the junction interface. $F$ is a function of the shape function $\frac{B(\alpha_{0})}{\alpha_{0}}.$

If matter is considered normal then $\sqrt{\eta}$ is usually interpreted as the speed of sound and is constrained to lie between $(0,1].$ However if matter is exotic then $\sqrt{\eta}$ can take any possible value \cite{povis}. Thus to determine the stability region we shall separate out the two cases, since depending on the parameters the matter surrounding the wormhole throat could be normal or exotic.  Using the stress-energy conditions one can easily find that NEC is obeyed if $\frac{B_{1}}{\zeta}\leq\frac{2}{3}\left(\frac{B(\zeta)}{\zeta}\right)$ and NEC is violated in case $\frac{B_{1}}{\zeta}>\frac{4}{3}\left(\frac{B(\zeta)}{\zeta}\right).$ Accordingly we shall analyse the region of stability as follows:
\begin{itemize}
\item[] If $0<\eta\leq 1$ then the function $\frac{B(\alpha_{0})}{\alpha_{0}}$ is constrained between $\frac{3}{2}\left(\frac{B_{1}}{\alpha_{0}}\right)\leq \frac{B(\alpha_{0})}{\alpha_{0}}<1.$ Recall that parameter $B_{1}$ has been interpreted as the location of the throat. 
\item[]If matter in the wormhole throat is exotic then $\eta$ can take any value and now $\frac{B(\alpha_{0})}{\alpha_{0}}$ is constrained between $0< \frac{B(\alpha_{0})}{\alpha_{0}}\leq\frac{3}{4}\left(\frac{B_{1}}{\alpha_{0}}\right),$ with $B_{1}$ having same interpretation.  
\end{itemize}
The meshed regions in Fig 1 and 2 depicts graphically the admissible stability region for both normal and exotic matter.

\begin{figure}[htb]
\centering
  \begin{tabular}{@{}ccc@{}}
\includegraphics[width= 0.5\linewidth]{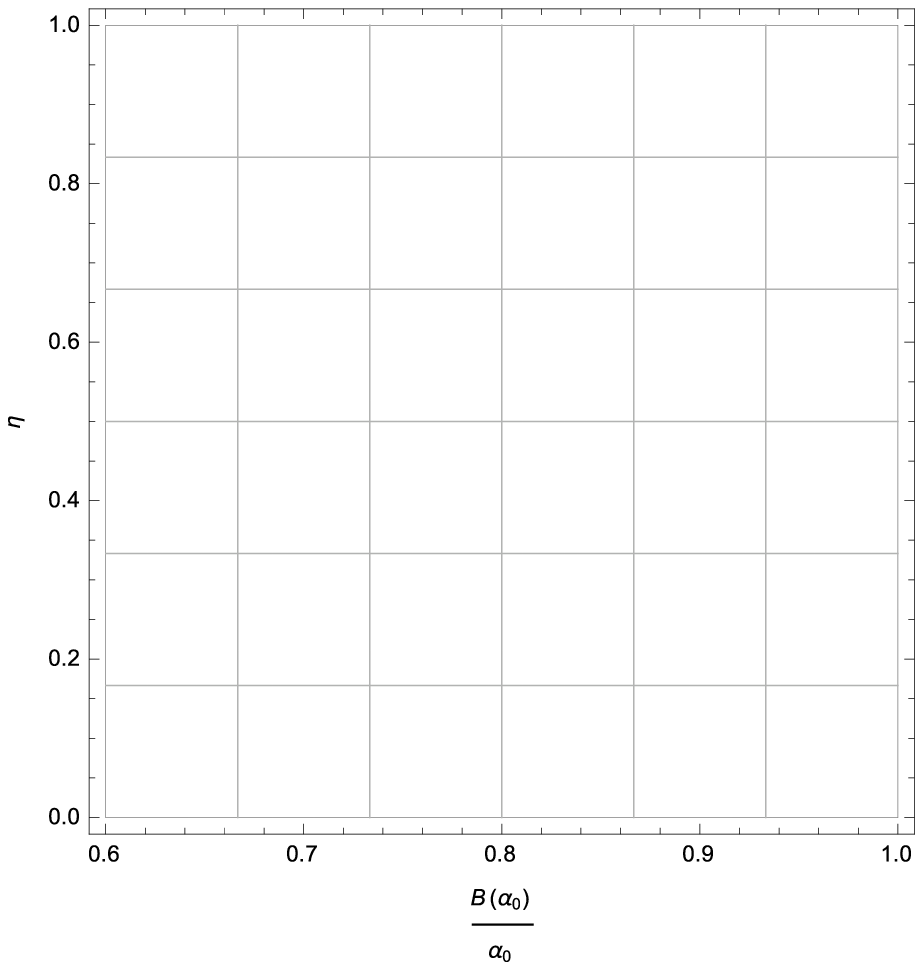}&
\includegraphics[width= 0.5\linewidth]{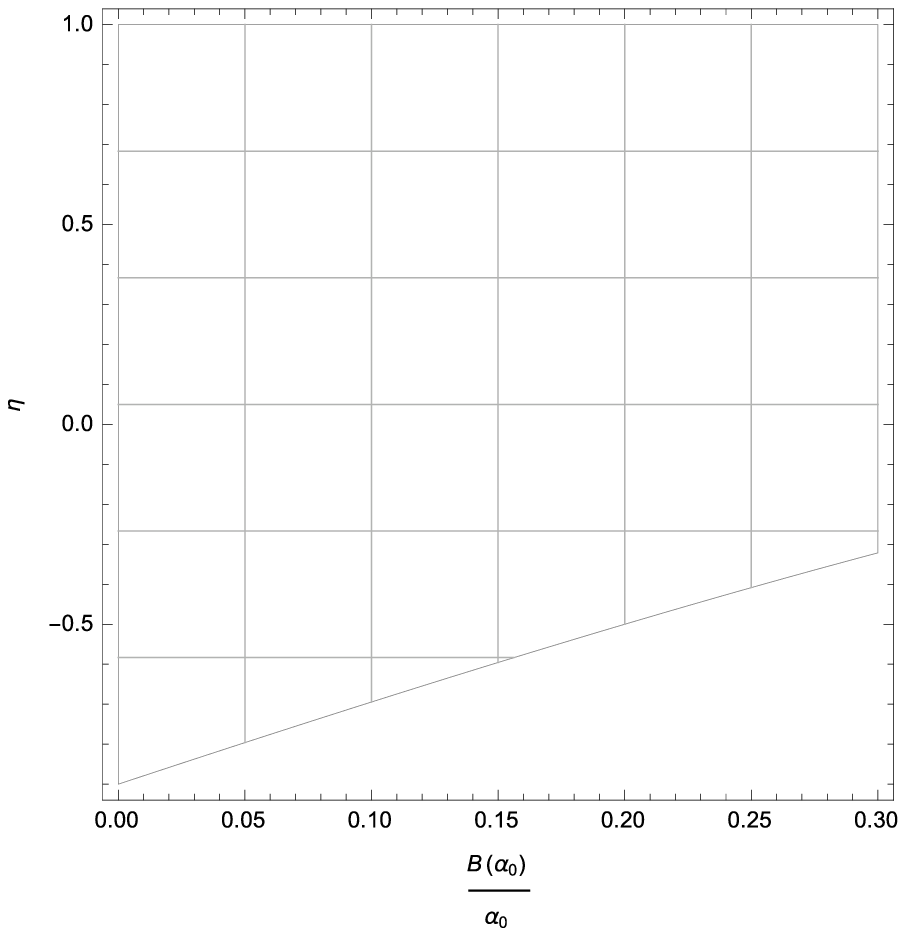}&
\end{tabular}
\caption{ The stability region for $\frac{B_{1}}{\alpha_{0}}=0.4$ where first panel corresponds to normal matter, while second panel corresponds to exotic matter.} 
\end{figure}

\begin{figure}[htb]
\centering
  \begin{tabular}{@{}ccc@{}}
\includegraphics[width= 0.5\linewidth]{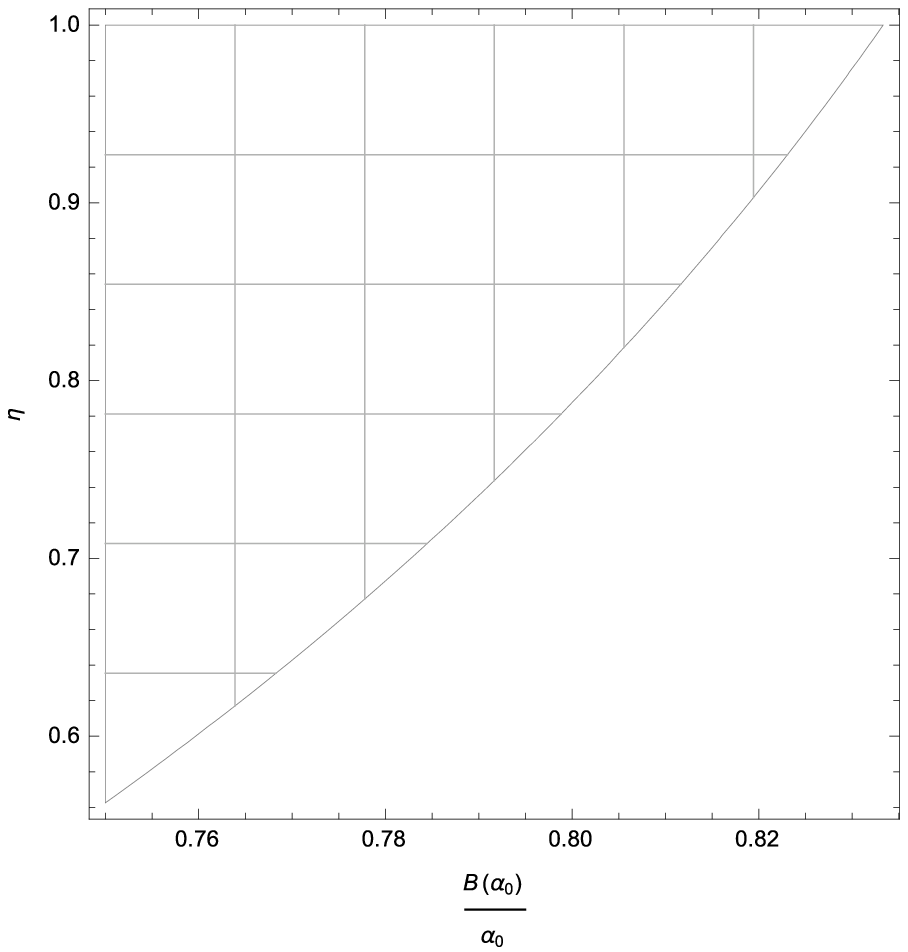}&
\includegraphics[width= 0.5\linewidth]{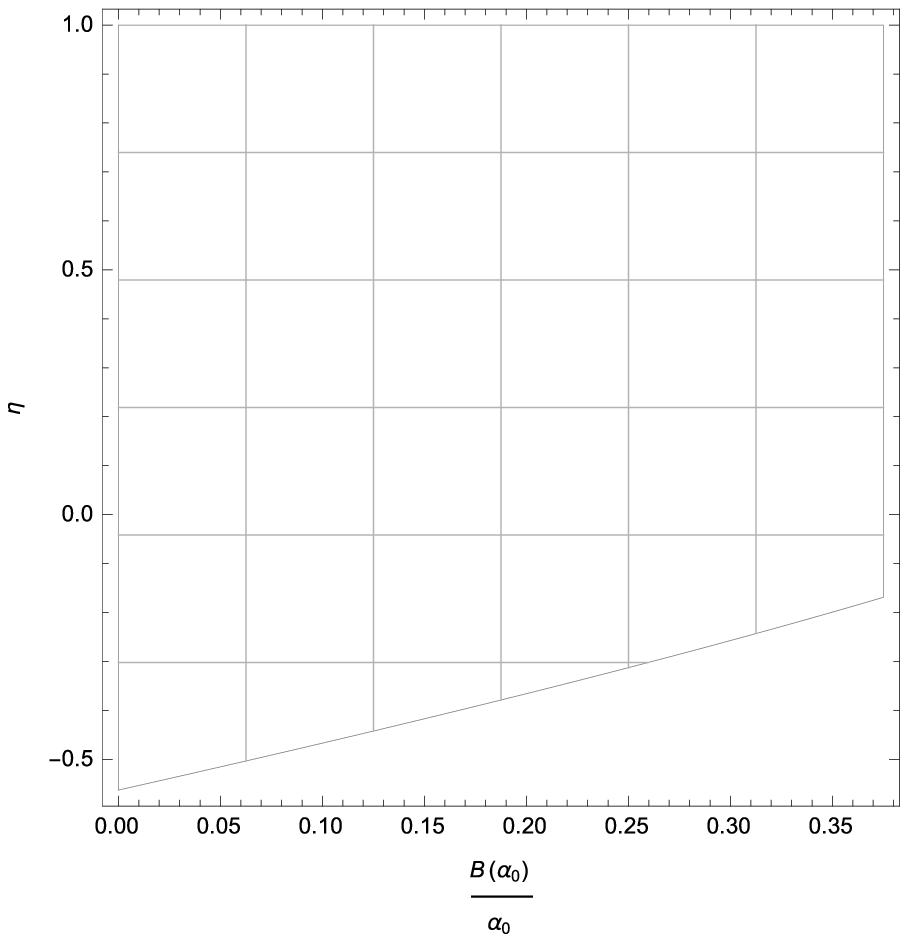}
\end{tabular}
\caption{ The stability region $\frac{B_{1}}{\alpha_{0}}=0.5$ where first panel corresponds to normal matter, while second panel corresponds to exotic matter.} 
\end{figure}

From the above descriptions we conclude that one can describe a complete cosmic scenario, along the entire time axis, that does not require any beginning or end. The scenario that we exhibited is ever existing and without any strong singularity. The inflationary epochs are realised at region after(before) emergence in the form of accelerated expansion(contraction). Further the otherwise isolated ends at $t=\pm\infty$ are found to be tunnelled through a wormhole, such that after final contraction one can get back to another expanding universe. Thus such an universe is ever existing and cyclical. Further the model adheres to basic principles of physics and presents a mathematically conceivable picture. 

We might as well emphasise that although based on phenomenological choices, the above model of cosmic evolution in some way combines all the existing models under a single umbrella and presents a complete picture of the universe, that resembles reality over a large period of time. Further this model presents inflationary process both as expanding and compressing mechanisms, complementing each other and hence making the story complete.

\end{document}